\documentclass[12pt,preprint]{aastex}
\begin{document}

\def\Msun{M$\odot$}

\newcommand{\vdag}{(v)^\dagger}
\newcommand{\myemail}{skywalker@galaxy.far.far.away}
\newcommand \lta {\mathrel{\vcenter
     {\hbox{$<$}\nointerlineskip\hbox{$\sim$}}}}
\newcommand \gta {\mathrel{\vcenter
     {\hbox{$>$}\nointerlineskip\hbox{$\sim$}}}}
\newcommand \m {M$_\odot$}
\newcommand \mm {M_\odot}
\newcommand\cms{cm~s$^{-1}$}
\newcommand\kms{km~s$^{-1}$}
\newcommand\gr{$\gamma$-ray}
\newcommand\grs{$\gamma$-rays}
\newcommand\grb{$\gamma$-ray burst}
\newcommand\grbs{$\gamma$-ray bursts}
\newcommand\degree{$^{\rm o}$}
\newcommand\etal{et al.}

\newcommand{\mn}{\ion{Mn}{2}}
\newcommand{\ca}{\ion{Ca}{2}}
\newcommand{\mg}{\ion{Mg}{2}}
\newcommand{\oi}{\ion{O}{1}}
\newcommand{\ci}{\ion{C}{1}}

\renewcommand{\thefootnote}{\fnsymbol{footnote}}
\newcommand{\mum}{$\mu$m}

\title{Pre-Maximum Spectropolarimetry of the Type Ia SN 2004dt
\footnote{Based on observations collected at the
European Southern Observatory, Chile (ESO Progr. No. 073.D-0565(A)).}}
 
\author{Lifan Wang$^{1,2}$, Dietrich Baade$^3$, 
Peter H\"oflich$^4$, J.~Craig~Wheeler$^4$, Koji~Kawabata$^5$, Alexei Khokhlov$^6$, 
Ken'ichi Nomoto$^7$, Ferdinando Patat$^3$}

\affil{$^1$Lawrence Berkeley National Laboratory, 1 Cyclotron Rd, Berkeley, CA 94712}

\affil{$^2$Purple Mountain Observatory, 2 West Beijing Road, Nanjing
210008, China}

\affil{$^3$European Southern Observatory,
    Karl-Schwarzschild-Strasse 2,
     D-85748 Garching, Germany}

\affil{$^4$Department of Astronomy and McDonald Observatory,
          The University of Texas at Austin,
          Austin,~TX~78712}

\affil{$^5$Astrophysical Science Center,
        Hiroshima University, 1-3-1 Kagamiyama, Higashi-Hiroshima, Hiroshima 739-8526, Japan}

\affil{$^6$Department of Astronomy, Department of Astronomy and Astrophysics, University of Chicago, 
Chicago, Illinois 60637, USA}

\affil{$^7$Department of Astronomy, University of Tokyo, Bunkyo-ku, Tokyo 113-0033, Japan}

\begin{abstract}

We report observations of SN~2004dt obtained with the 
Very Large Telescope of the European Southern Observatory on 
August 13.30, 2004 when the supernova was more than a week before 
optical maximum. SN~2004dt showed strong lines of \ion{O}{1}, \ion{Mg}{2}, 
\ion{Si}{2}, and \ion{Ca}{2} 
with typical velocities of absorption minimum around 17,000 \kms. The 
line profiles show material moving at velocities as high as 25,000 
\kms\ in these lines. The observations also reveal absorption lines 
from \ion{S}{2} and \ion{Si}{3} with a velocity of only 11,000 \kms. 
The highest velocity in the \ion{S}{2} features can be traced to about  
15,000 \kms, much lower than those of O, Mg, Si, and Ca. 
SN~2004dt has a polarization spectrum unlike any previously observed. 
The variation of the polarization across some \ion{Si}{2} lines 
approaches 2\%, making SN~2004dt the most highly polarized SN~Ia ever 
observed.  In contrast, the strong line of O I at 777.4 nm shows little or 
no polarization signature. The degree of polarization points to a 
silicon layer with substantial departure from spherical symmetry.  
A geometry that would account for the observations is one in which
the distribution of oxygen is essentially spherically symmetric, but 
with protrusions of intermediate-mass elements 
within the oxygen rich region.

\end{abstract}
 
\keywords{supernovae -- cosmology -- spectropolarimetry}

\section{Introduction}

As for core-collapse supernovae, high-quality spectropolarimetry of
Type Ia supernovae (SNe~Ia) is revealing a complex array of
behavior that has opened a new window of exploration of this
class of explosions (Wang et al. 1996; Wang, Wheeler, H\"oflich 1997; 
Howell et al.  2001; Wang et al. 2003a; Wang et al. 2004). 
While there are indications that the
degree of polarization of core-collapse supernovae increases with time
after explosion (Wang et al. 2001, 2002; Leonard et al. 2001), 
the time evolution of Type Ia supernovae seems to be rather different
(Wang et al. 2003a).

Polarimetry of SN~Ia shows that the degree of polarization of 
normal SN~Ia is $\lesssim$ 0.3\%, which is much lower than 
the typical polarization of core-collapse supernovae (Wang et al. 1996).  
Spectropolarimetry of the normally-bright 
SN~1996X at optical maximum tentatively showed intrinsic polarization 
of about 0.3\% which implies geometrical asphericities of about 10\%
(Wang et al. 1997).  Hydrodynamic instabilities just beyond the 
layers of complete thermonuclear equilibrium may have contributed 
to the polarization signal from SN~1996X. 
Howell et al (2001) showed that SN~1999by, a sub-luminous SN~Ia, 
was polarized at a level of 0.7\% in the red continuum. 

The observations of the normally-bright SN~2001el revealed the geometric 
structure of an SN~Ia in different chemical layers. The degree 
of polarization showed significant time evolution (Wang et al. 2003a). 
An exceedingly strong \ion{Ca}{2} feature at velocity of 20,000 - 26,000 
\kms\ was highly polarized. This feature showed that the outermost 
layers of the supernova ejecta had a highly aspherical geometry 
(Wang et al. 2003a; Kasen et al. 2003).  Deeper inside the ejecta, 
the silicon-rich layer also displayed asphericity as evidenced 
by the detection of polarization of $\sim 0.3\%$ for 
the \ion{Si}{2} 635.5 nm line. When the photosphere receded deep 
inside the iron-rich layers, the polarization diminished and 
the supernova appeared spherical. Wang et al. (2003a) listed 
several possible origins of this asymmetric high-velocity 
material: matter from a disk or companion star that is swept 
up by the supernova ejecta; newly-synthesized material produced by 
thermonuclear reactions in the high velocity collision between supernova 
ejecta and the surroundings; 
and a clump of matter ejected from the intrinsic thermonuclear 
combustion process. Gerardy et al. (2004) examined the 
high-velocity \ion{Ca}{2} feature in SN~2003du (for which there 
was no spectropolarimetry) and concluded that the strength 
of the \ion{Ca}{2} in that event was consistent with solar abundance 
of calcium in a hydrogen plasma (that the bulk of the matter was 
helium or carbon/oxygen cannot be ruled out) that was swept up from 
distances consistent with the expected separation of the companion star. 

We present in this paper observations of SN~2004dt.  This event
showed exceptionally strong and broad lines of \ion{Si}{2},
and \ion{O}{1} and revealed the strongest polarization signal we
have yet measured for a SN~Ia. The discovery and pre-maximum observations are
described in \S 2, the spectral and spectropolarimetry
features and their implications for ejecta geometry are
discussed in \S 3. 
The inferred ejecta geometry is given in \S 4. The observations
of SN~2004dt are compared to other SN~Ia in \S 5. Comments on 
theoretical models are given in \S 6, based on the observational 
results of SN~2004dt. We give in \S 7 some discussions on the
impact of asymmetry on the use of SN~Ia as cosmological probes.
Finally, in \S 8, we provide a brief summary and some 
general discussions of the most important conclusions we 
have derived from spectropolarimetry.
 
\section{Discovery and observations}

\subsection{Discovery}

SN 2004dt in NGC 799 was discovered on an unfiltered image 
taken on August 11.48 at a magnitude of about 16.1 
(Moore \& Li 2004). Salvo, Schmidt \&  Wood (2004) reported 
that a spectrum taken on August 12.72 UT showed SN~2004dt to 
be a young SN~Ia a few days before maximum light. The photospheric 
expansion velocity deduced from the absorption minimum of 
\ion{Si}{2} 635.5 nm was about 16,500 \kms, adopting the 
recession velocity of 5915 \kms\ reported by the NED for the host galaxy NGC 799.  
Patat, Pignata, and Benetti (2004) reported that a spectrum 
obtained on Aug. 13.17 UT showed some unusual features. They 
noted  that the \ion{O}{1} 777.4 nm transition was particularly 
pronounced, with a broad symmetric P-Cygni profile and also
had an expansion velocity of about 16,800 \kms.  They also 
report that the \ion{Ca}{2} H \& K line gave 22,200 \kms\ (we 
think this feature is mostly due to a \ion{Si}{2} line, see \S3.2). 
Patat, Pignata, and Benetti (2004) suggested that the high velocities of the lines indicated a 
very early phase.  They noted that other intense lines visible 
in the spectrum were from Fe III and Si III.

\subsection{Observations and data reduction}

We observed SN~2004dt with the ESO-VLT on 2004 August 13.30 UT 
with the FORS1 instrument in polarization mode. From preliminary
light curves of Li et al. (2004, private communication), this corresponds
to 6-8 days before optical maximum.
Four exposures of 1200 second durations were taken with the waveplate at 
position angles of 0, 45, 22.5, and 67.5 degrees. Flux and 
polarization standard stars were observed to calibrate the 
flux levels and polarization position angles.  The data were 
reduced in a way similar to that described in Wang et al. (2003a).

\subsubsection{Spectrophotometry}
\label{specphot}

The observations consist of multiple exposures with the airmass varying 
from 1.02 to about 1.20. The slit size was 1 arcsec and was aligned 
with the paralactic angle. The seeing was below 1 arcsec 
which allows for approximate spectrophotometry calibrations.
Fig.~\ref{spec} (top panel) shows the total flux spectrum.
Starting in the blue, there is no sharp drop of flux blue ward of the 
strong \ion{Ca}{2} H \& K absorption as commonly observed in 
normal SN~Ia at optical maximum. The \ion{Ca}{2} H \& K 
feature appears very strong although, as we argue in \S3.2, it 
should probably be identified as \ion{Si}{2} 385.9 nm. The strong 
line of \ion{Si}{2} 635.5 nm shows the typical blue-shifted 
absorption that characterizes SN~Ia at 600 nm, but it is 
especially strong, broad, and fast moving.  The minimum corresponds to a 
velocity of about 17,000 \kms, and the blue wing extends to 
about 25,000 \kms. We also observe a strong line of 
\ion{O}{1} 777.4 nm as pointed out by Patat, Pignata, and Benetti (2004).
This line also showed an absorption minimum at about 17,000 \kms\
with the blue wing again extending to about 25,000 \kms. To
first order, the Si and O occupy the same velocity space. 
Our data show two components of the \ion{Ca}{2} IR triplet, 
a broad absorption at 820 nm and a ``notch" indicating a separate,
high-velocity component at about 800 nm. These represent 
velocities of $\sim$ 16,000 and 23,000 \kms, respectively. 
The lower velocity component corresponds approximately to the
velocity minimum of \ion{Si}{2} and \ion{O}{1}. The higher
velocity component seems kinematically distinct as in 
SN~2001el (Wang et al. 2003a) and SN~2003du (Gerardy et al. 2004)
and to correspond roughly to the blue wing of the \ion{Si}{2} 
and \ion{O}{1} lines. Despite the likely two-component nature, 
the highest velocity, again, can be traced
to 25,000 \kms\ for the \ion{Ca}{2} IR triplet, similar to the other
high-velocity lines.

The strong absorption feature at 421 nm is due to 
\ion{Mg}{2} 448.1 nm, blended with \ion{Fe}{3} and \ion{Si}{3} lines. 
If this feature is predominantly \ion{Mg}{2}, then the absorption minimum 
of the \ion{Mg}{2} line is blueshifted by 17,000 \kms, consistent with
the lines of \ion{Si}{2}, \ion{Ca}{2}, and \ion{O}{2}.
The strength of the 421 nm feature is comparable to the \ion{O}{1} 777.4 nm
line, as measured by the ratio of line minimum to local continuum.

The absorption minima of the \ion{S}{2} ``W" lines are measured to be at
523.5 and 543.1 nm which implies a velocity of only 11,000 \kms.
This is significantly lower than the velocity of the \ion{Si}{2} lines. 
Weak lines at 439.5, 454.2, and 464.0 nm, tentatively identified 
with \ion{Si}{3} 456.0 nm, \ion{S}{2} 471.6 nm, and \ion{S}{2} 481.6 nm, 
respectively, also show the much lower velocity of 11,000 \kms.
The presence of \ion{Si}{3} 456.0 nm at 11,000 \kms, not at 17,000 \kms\ 
shows that doubly ionized Si exists only close to the central region.
Silicon is the only element present in two different ionization
stages. They share the emerging bimodal distribution of radial
velocities: Singly ionized silicon has a bulk velocity of 17,000 \kms
whereas \ion{Si}{3} expands with 11,000 \kms.

Table~1 gives a schematic overview of the kinematic properties of the
main ions observed.

\subsubsection{Spectropolarimetry}
\label{specpol}
\begin{figure}
\figurenum{1}
\epsscale{.8}
\plotone{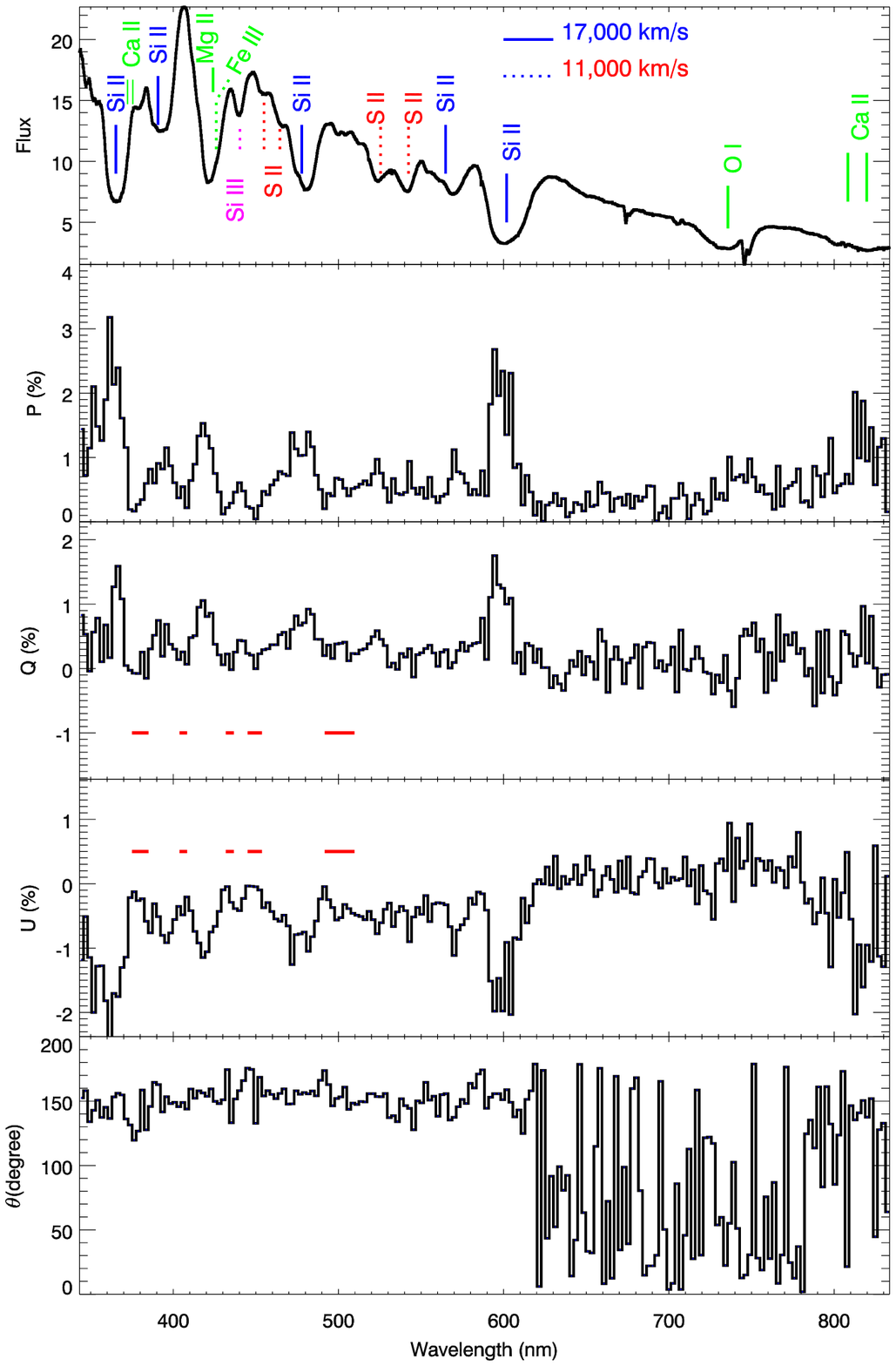}
\caption{Spectroscopy of SN~2004dt de-redshifted assuming a recession
velocity of 5915 \kms\ of the host galaxy. The four panels give the total
flux spectrum, the polarization, P, the reduced Stokes vectors, Q and U,
and the polarization angle $\theta$. Typical spectral features are marked 
on the top panel. The lines are produced by material moving at
velocities of 17,000 \kms\ (marked by solid lines) and 11,000 \kms\  
(marked by dotted lines; see Fig. 3 for a more complete description) in the
rest frame of the supernova. Lines from different elements are marked
with different colors (gray scales). The 
panels for polarization data are binned to
25 \AA. The wavelength regions used to determine interstellar polarization
are marked by horizontal lines on the panels for Q and U.
Stronger lines tend to be associated with larger degrees of polarization
except for the \ion{O}{1} 777.4 nm line.
}
\label{spec}
\end{figure}

The second, third, and fourth panels of Fig.~\ref{spec} show the
polarization, $P$, and the reduced Stokes vectors, $U$, and $Q$. The
degree of polarization is corrected for biases due to observational
noise using the equations derived in Wang, Wheeler, \& H\"oflich
(1997).  
The $Q$ vector, in particular, shows a total excursion of 
$\sim$ 2 percent with sharp wavelength dependence, indicating
strong, intrinsic polarization. The bottom panel of Fig.~\ref{spec}
gives the polarization position angle. It is nearly constant from 330 nm
to 630 nm, but then shows a strong change with possibly
significant excursions at wavelengths longer than 630 nm. This change
could be due to a combination of polarization by interstellar dust
and the intrinsic polarization by the supernova ejecta.
Note that the polarization position angle is periodic with respect
to the position angle on the sky with a period of 180\degree.

\begin{figure}
\figurenum{2}
\epsscale{.9}
\plotone{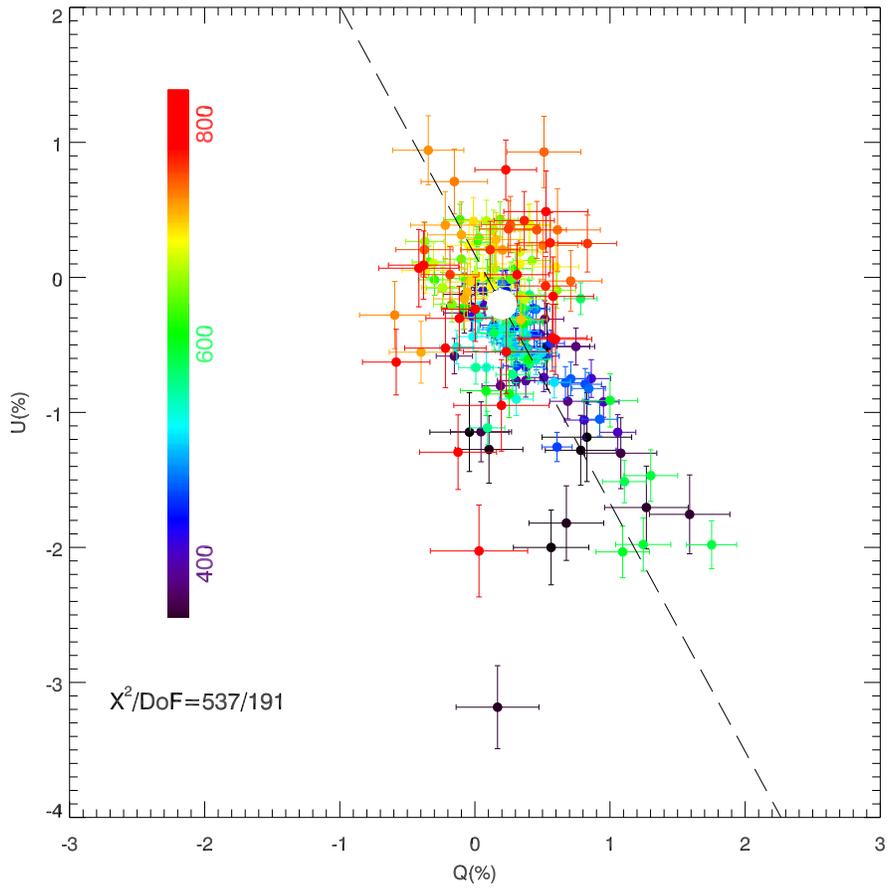}
\caption{Spectropolarimetry of SN~2004dt displayed in the plane of
the reduced Stokes vectors Q and U. Note the existence of a 
dominant axis represented by the solid line. The 1-$\sigma$ 
location of the polarization due to interstellar dust is shown 
as an open circle. The data points are binned to 25 \AA.
}
\label{qu}
\end{figure}

Fig.~\ref{qu} gives the data in the
Q/U plane where each point represents the Stokes parameters
rebinned to 25 \AA\ intervals. The bin size was chosen to be
larger than the spectral resolution to minimize correlations
among neighboring lines. The total excursion is nearly
3 percent. The scatter of points is predominantly due to
line features.  The data points span a range of
about 1.5 percent in Q and more than 2 percent in U.  
Fig.~\ref{qu} shows that the bulk of the points fall along 
the line that represents the ``dominant axis" of the
geometry (Wang et al. 2001), determined by making a principle
component decomposition of the data points on the Q/U plane. 
The scatter orthogonal to this axis suggests that there are 
smaller but real deviations in the geometry from this 
dominant axis that may also constrain the physics of the explosion. 

\section{The distinguishing observational features}

\subsection{The spectral lines and profiles}

The colors synthesized from the spectrum give $U - B \sim$ -0.6.
This blue color (as verified by the strong blue flux in Fig.~\ref{spec}),
suggests that this event, if approximately ``normal," is close to
or before $U$ maximum. This may mean that at this phase, 
SN~2004dt has not yet been subject to the strong \ion{Fe}{2} 
absorption leading to 
the characteristic UV deficit of normal SN~Ia near maximum light 
(Wheeler \& Harkness 1992; Kirshner et al. 1993). 

\begin{figure}
\figurenum{3}
\epsscale{.8}
\plotone{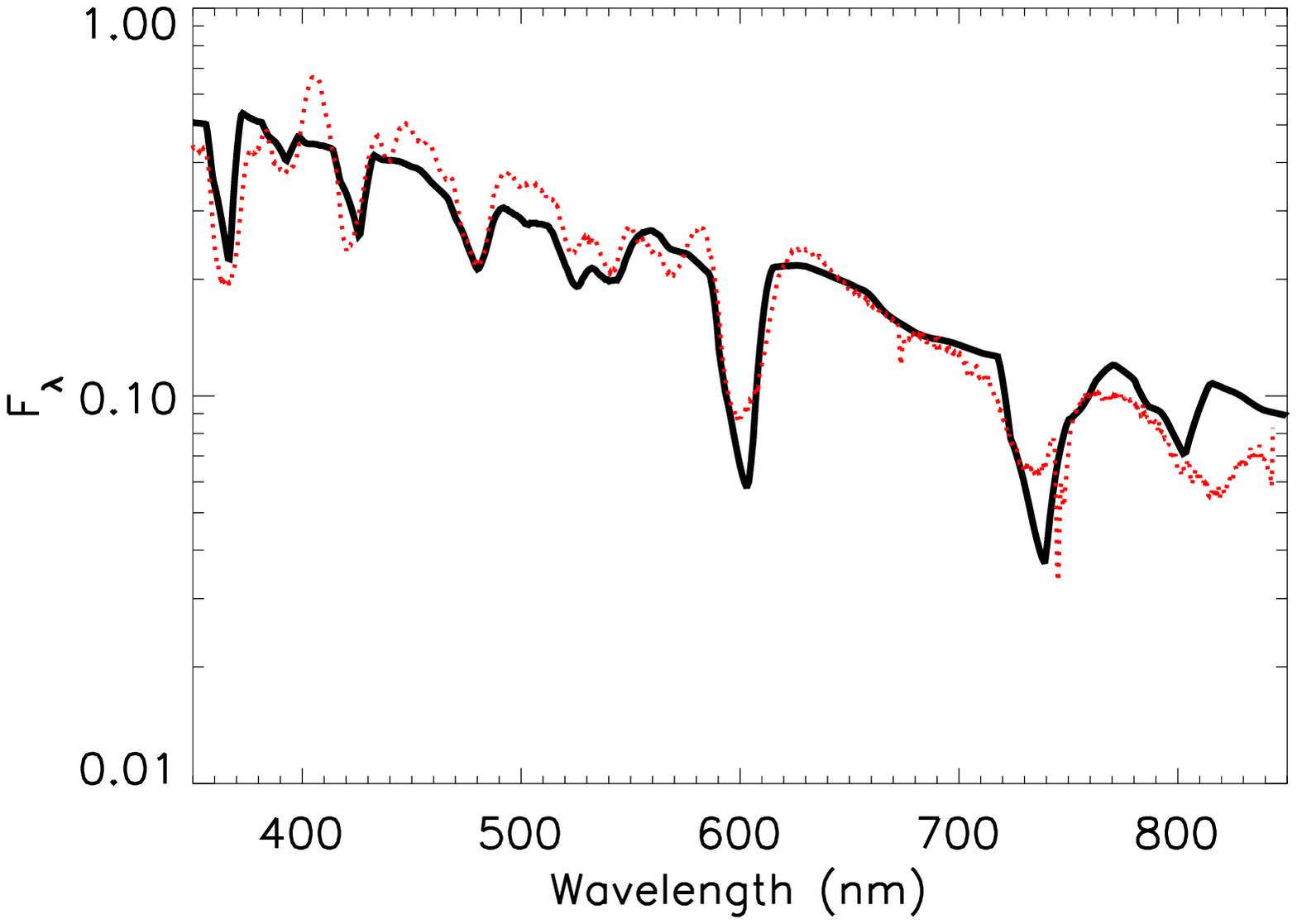}

\caption{Line identifications in the observed spectrum using the SYNOW
code. The only ions contributing to the model spectrum (solid line)
are \ion{Si}{2}, \ion{Mg}{2}, \ion{O}{1}, and \ion{S}{2}. Other relevant
fit parameters are: The ions are distributed in a shell from 16,000 \kms\ 
to 25,000 \kms, with the exception of the \ion{S}{2} which is distributed
from 11,000 \kms\ to 18,000 \kms; the location of the photosphere is 
assumed to be at 11,000 \kms\ with the temperature fixed to 11,000 K; 
the optical depth is assumed to decrease outwards with an exponential law
with an e-folding velocity of 5 \kms; and the optical depths used in
SYNOW are 2.0, 0.5, 0.5, and 1.0 for \ion{Si}{2}, \ion{Mg}{2}, \ion{S}{2},
and \ion{O}{1}, respectively. The unusual appearance of the observed
\ion{Si}{2} profiles are likely due to a combination of the 
high velocity and overabundance of \ion{Si}{2}.
}
\label{synow}
\end{figure}

The strong lines of \ion{Si}{2} 635.5 nm and \ion{O}{1} 777.4 nm 
in the total flux spectrum suggest that there is substantial ejected 
silicon and oxygen, moving at velocities approaching 25,000 \kms. This 
is consistent with a significant amount of incompletely 
burned matter. The lack of a UV deficit suggests that 
there is not much iron in the photosphere at the phase of these 
observations.
To investigate further the possibility of an iron deficit, 
we have performed line identifications using 
the SYNOW code (Fisher et al. 1999).  Assuming a composition with only
\ion{Si}{2}, \ion{S}{2}, \ion{O}{1}, and \ion{Mg}{2}, we found that
the observed spectral features are reasonably well 
explained without requiring iron. What distinguishes 
SN~2004dt from other Type Ia supernovae appears to be that 
a larger than usual amount of silicon is required to explain the
deep spectral features from \ion{Si}{2}. The SYNOW line identification
is shown in Fig.~\ref{synow}.
The absence of Fe features is another striking 
property of the spectrum of SN~2004dt at this epoch.

The presence of lines at a velocity of 11,000 \kms\ suggests that 
even at this early phase, the ejecta is not Thompson opaque at 
this lower velocity. The velocity of the \ion{Si}{3} 455.3, 456.8, 
and 457.5 blend is only 11,000 \kms, implying that doubly-ionized 
silicon exists mainly 
in this lower-velocity region. The low velocity \ion{S}{2} 
line suggests that S is more concentrated in the low-velocity region
than Ca, O, Mg, and Si.

\subsection{The interstellar polarization toward SN~2004dt} 

From the strong modulation of the raw polarization shown in 
Fig.~\ref{spec} it is immediately evident that the intrinsic 
linear polarization of SN~2004dt is exceptionally large. 
The vectorial nature of
polarization can lead to confusing conclusions if the
interstellar polarization (ISP) is not properly determined and
subtracted; unfortunately, the determination of ISP 
inevitably requires some model
assumptions. We have tried to avoid assumptions that are tied to the
models of SN Ia explosions.

The central assumption made here is that the ISP dominates where the
continuum polarization of the supernova is the lowest. The effective
continuum polarization probably gets decreased by overlapping spectral
lines. Since there are many more such lines in the blue portion of the
spectrum than in the red 
portion (Wang et al.\ 2001; Howell et al.\ 2001; Wang et al.\ 2003a), 
we have selected the regions marked in Fig.~\ref{spec} 
to represent the minimal intrinsic polarization; they avoid 
significant discrete absorption features.

It is seen in Fig.~\ref{qu} that the deduced ISP corresponds to $0.2\pm0.1\%$
in $Q$ and $-0.2\pm0.1\%$ in $U$. This low value is consistent with
the absence of an interstellar Na I D line (\S 3.6) if the
gas-to-dust ratio of the host galaxy is normal.

In Fig.~\ref{qu}, the data points used for the determination of the ISP are
embedded in a large cloud of other low-polarization wavelength bins.
If, following Wang et al.\ (2001), a principle-component
decomposition technique is used to define the dominant and the
orthogonal polarization axes, the ISP will, therefore, be close to the
dominant axes.  As Wang et al.\ (2001) have shown, some minor error in
the determination of the ISP does not grossly distort a qualitative
interpretation if the true ISP is close to the dominant axis.

A similar constellation of data points
prevailed in SN~2001el (Wang et al. 2003a).  The 
key corroborating observation was that at later phases the polarization 
of the continuum as well as of spectral features retreated along the 
dominant axis and, starting a week after maximum, remained within 
a similar cloud of low-polarization wavelength bins.  A preliminary 
analysis of later-epoch data of SN~2004dt shows a similar behavior.

The resulting ISP-corrected decomposition into dominant and orthogonal
axis is presented in Fig.~\ref{polprofiles}.

\subsection{The intrinsic polarization of SN 2004dt}

Ultimately, the choice of interstellar polarization must depend on the
totality of observations and a self-consistent picture of the physics
of the explosion and radiative transfer through the complex ejecta.
Important details of the interpretation of the data may yet depend on the
particular placement of the ISP. For this paper we have attempted to
restrict our discussion to features that are relatively independent
of the choice of the ISP.

SN~2004dt shows a polarization spectrum unlike any SN~Ia yet
observed. Compared to the well-observed SN~2001el, the \ion{Si}{2} lines
are especially prominent in the polarization spectrum, but
the lines of the Ca IR triplet are less so.

With the above choice of the interstellar polarization, we see from
Fig.~\ref{polprofiles} that the absorption minima of the P-Cygni profiles
correspond in general to the peaks of the spectropolarization
profiles.  Along the dominant axis polarization spectrum, the
polarization in the range 620 - 720 nm is around -0.3\% for the choice
of interstellar polarization adopted in Fig.~\ref{polprofiles}. 
This wavelength
range shows no strong spectral lines and it is likely that continuum
polarization by Thompson scattering dominates this wavelength region.

The high quality data we obtained allow for a detailed comparison of
the spectral and spectropolarimetry profiles of several distinct
lines, as shown in Fig.~\ref{lineprofiles}. Spectropolarimetry allows 
us to revisit the issue of line identification as discussed already
in \S3.1.

\subsubsection{\ion{Si}{2} lines}

\begin{figure}
\figurenum{4}
\epsscale{.8}
\plotone{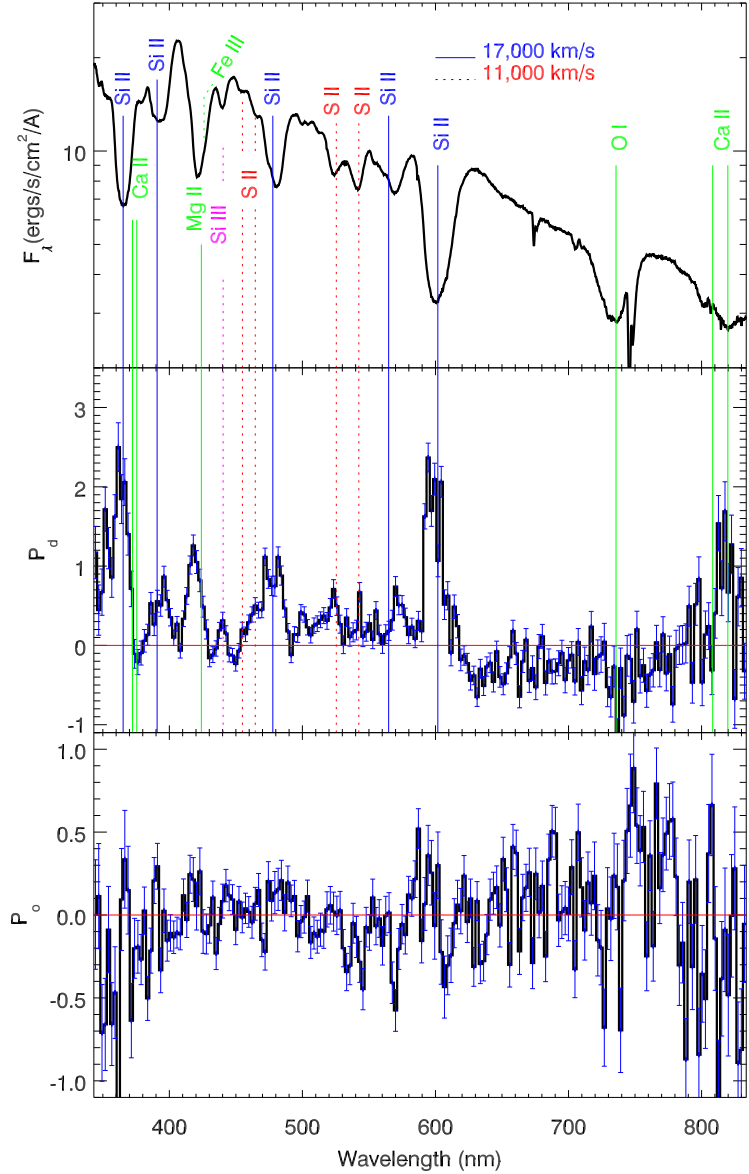}
\caption{
The top panel gives the total
flux spectra; the next two panels give the polarization spectrum projected
onto the dominant axis and the orthogonal axis, respectively. The solid
vertical lines mark the positions of the lines of \ion{Si}{2}, \ion{Mg}{2},
\ion{Ca}{2}, and \ion{O}{1} which are typically blue shifted by
17,000 \kms. The vertical dash lines mark the positions of the lines
of \ion{S}{2}, \ion{Si}{3}, and \ion{Fe}{3}, at a blue shift of 11,000 \kms.
All the polarized features except \ion{O}{1} line share the same dominant 
axis, suggesting that the geometrical structures of \ion{Si}{2}, \ion{Mg}{2},
and \ion{Ca}{2} are the same. No polarization is detected across the
\ion{O}{1} line. Weaker lines of \ion{Si}{3}, and \ion{S}{2} show also
polarization along the dominant axis. The data are binned to 25 \AA.
Note that the polarization features at 747.0 nm may not be reliable because of
telluric absorption.
}
\label{polprofiles}
\end{figure}

\begin{figure}
\figurenum{5}
\epsscale{.8}
\plotone{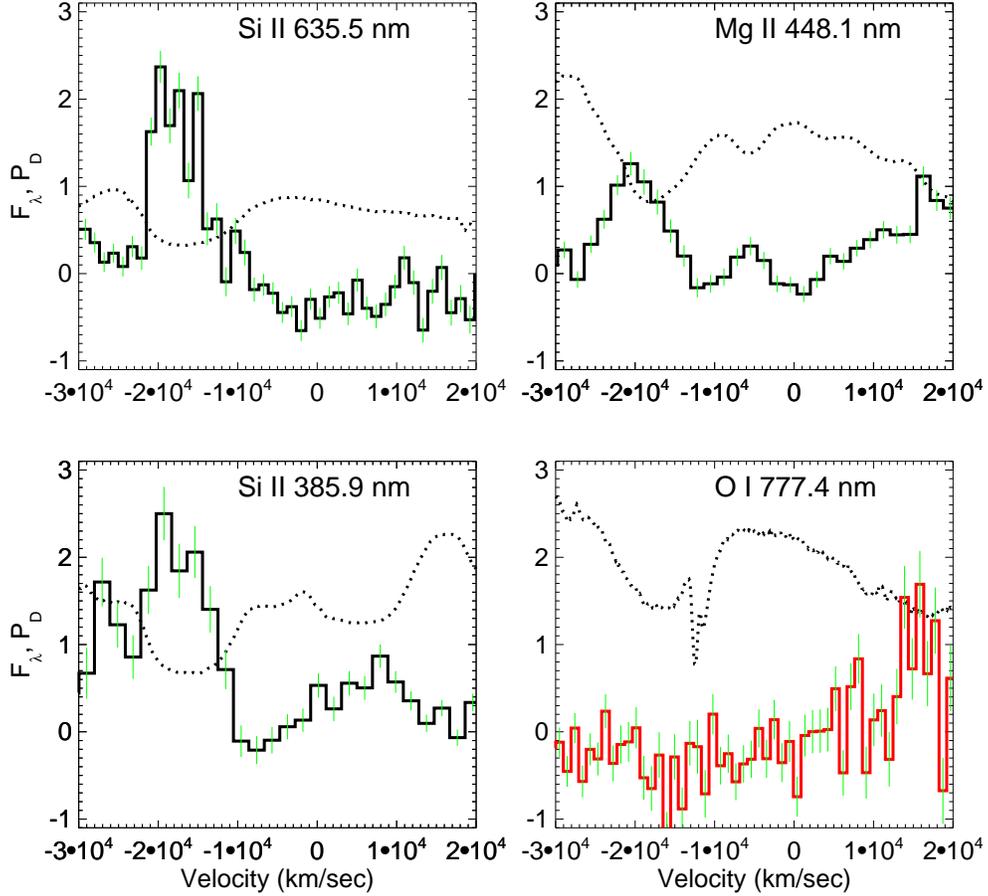}
\caption{
The absorption features in the total flux spectrum and the
polarization profiles along the dominant axis are shown as a
function of velocity for the lines of \ion{Si}{2} 635.5 nm (a, top left),  
\ion{Mg}{2} 447.1 nm (b, top right),  \ion{O}{1} 777.4 nm (c, lower right), 
and \ion{Si}{2} 385.9 nm (d, lower left). The dotted lines show the 
P-Cygni profiles. The solid histograms shows the spectropolarimetry
profiles. 
Note that the polarization is detected in all lines except that of
\ion{O}{1}. Note also that the emission component of the 
P-Cygni profiles are all blueshifted by about 4,000 \kms.
}
\label{lineprofiles}
\end{figure}

One of the strongest polarization features in Fig.~\ref{polprofiles} and 
Fig.~\ref{lineprofiles} is the spike at
600 nm which corresponds to the \ion{Si}{2} 635.5 nm line. The strength
of this feature, $\sim$ 2\%, is unprecendented in the observed
polarization spectra of SN~Ia. 
There is also a strong spike in the polarization at 365.0 nm, the location
of the ``Ca H \& K" line. We suspect this is due primarily to
\ion{Si}{2} 385.0 nm that is only slightly contaminated by the 
Ca lines. Note
that the amplitude of this polarization feature, $\sim$ 2\%, is nearly
identical to that of the \ion{Si}{2} 635.5 nm line.  No other features
in the polarization spectrum show this level of polarization. 
The spectral and spectropolarimetry
profiles of \ion{Si}{2}$\lambda\lambda$385.9, 635.5 nm lines are shown in
Fig.~\ref{lineprofiles}(a) and (c), respectively.
The polarizations are the highest at the absorption minima of these
two features. The spectral
and spectropolarization profiles of the two lines are remarkably
similar. The peaks of polarization occur at a blueshift of 
about 20,000 \kms for both lines then drop sharply at higher
velocity. This strengthens
our identification that they are from the same ionic species. 
The dominant-axis spectrum shows another feature at 390 nm that we
attribute to \ion{Si}{2} 413.0 nm.  Note, however, that the same
feature was tentatively identified as
\ion{C}{2} 426.7 nm with an expansion velocity of about 24,100 \kms\
by Patat et al. (2004).
Our identification with \ion{Si}{2} would put the line at the same
velocity of about 17,000 \kms\ as the other \ion{Si}{2} lines.

The line at 480 nm can be due to the blend of \ion{Si}{2} 504.1, 505.6 nm, and
\ion{Fe}{2} 491.3, 501.8, 516.9 nm. As can be seen in Fig.~\ref{polprofiles},
the contributions from \ion{Fe}{2} must be small as evidenced by the
absence of the strongest \ion{Fe}{2} line in this wavelength region,
\ion{Fe}{2} 516.9 nm, in the total flux spectrum.
The small feature at 570 nm is likely to be due to \ion{Si}{2} 595.8, 597.9 nm.
To the red of the \ion{Si}{2} 635.5 nm line, the polarization
dips and is then constant to about 720 nm.
With these identifications, all the
Si II lines have velocities from 15,000 to 17,000 \kms.

The lines of \ion{Si}{2} 413 nm, \ion{Mg}{2} 447.1 nm, and
\ion{Si}{2} 504.1, 505.6 nm all have polarization $\sim$ 1\% in
Fig.~\ref{polprofiles}, significantly less than the strongest polarized lines
of \ion{Si}{2}, but suggesting some commonality in their geometry.
All these lines have similar dominant polarization position angles.

\subsubsection{\ion{Si}{3} and \ion{S}{2} lines}

Weaker lines from \ion{Si}{3}, and \ion{S}{2} are
also polarized, at a level of about 0.3\%, and with polarization
position angles similar to those of the \ion{Si}{2} lines. 
The blueshifts of the absorption minima for these lines are 
around 11,000 \kms which is significantly lower than the strong
\ion{Si}{2} lines. 

\subsubsection{\ion{Mg}{2} 420.0 nm}

The polarization feature at 420 nm is mostly due to \ion{Mg}{2} 447.1 nm.
The profiles are shown in Fig.~\ref{lineprofiles}(b) for this line.
Like the polarizations of the \ion{Si}{2} 635.5 nm and \ion{Si}{2} 385.9 nm lines, 
the polarization of the \ion{Mg}{2} 447.1 nm line peaks also
at a blueshift of around 20,000 \kms, but the polarization does not
exhibit a sharp drop at higher velocities and appears to extend
to even higher velocities than those of the \ion{Si}{2} lines.

\subsubsection{\ion{Ca}{2} lines}

The polarized spectral feature at 820 nm
is clearly due to the \ion{Ca}{2} IR triplet;
the absorption minimum
of the \ion{Ca}{2} feature shows a velocity of 16,000 \kms, consistent
with that of the \ion{Si}{2} lines.
Note that, unlike SN~2001el which showed a highly polarizing
detached \ion{Ca}{2} shell at a velocity of 22,000 \kms,
the high-velocity component of the \ion{Ca}{2} IR triplet
of SN~2004dt with a similar velocity is not particularly strong nor
does it have an especially high polarization. SN~2001el and
SN~2004dt are distinctly different in this regard.
The lower velocity component of the \ion{Ca}{2} IR triplet in SN~2004dt
shows a polarization spike of about 0.7-2\%, comparable to the \ion{Mg}{2} line.
The higher velocity component shows a spike of about 0.3\%,
comparable to the \ion{S}{2} and \ion{Si}{3} lines. These
polarized components of \ion{Ca}{2} seem to share approximately the
same polarization angle as all the other polarized lines, $\sim$ 150\degree.

At the same velocity as the \ion{Ca}{2} IR triplet, only a weak dip is observed
for the \ion{Ca}{2}~H\&K lines. The weak \ion{Ca}{2}~H\&K lines do not produce 
a prominent polarized feature.

\subsubsection{\ion{O}{1} 777.4 nm line}

Despite the very strong spectral line of \ion{O}{1} 777.4 nm 
in the total flux spectrum, there is little sign of that feature
in the polarization spectrum. The spectral and spectropolarimetry
profiles of this line are shown in Fig.~\ref{lineprofiles}(c). This 
low polarization is in sharp contrasts to all the other
strong lines from \ion{Mg}{2}, \ion{Si}{2}, and \ion{Ca}{2}. 

\subsection{Summary of observed features}

The rather diverse spectral information extracted from our data can be
summarized as follows (see also Table~1):  (1) Lines of
\ion{Mg}{2}, \ion{O}{1}, \ion{Ca}{2}, and
\ion{Si}{2} are at the highest velocities, their absorption minima
are typically blue shifted by 17,000 \kms, and the lines can be traced
to a velocity of around 25,000-28,000 \kms; (2) The \ion{S}{2} lines have
absorption minima at only 11,000 \kms; the highest velocity of these
features is less than 15,000 \kms; (3) The absorption minimum of the
\ion{Si}{3} 455.3, 456.8, and 457.5 blend is blue shifted by 11,000
\kms; (4) All the high-velocity features except that from \ion{O}{1}
777.4 nm are significantly polarized, at levels from 0.7\% to 2.2\%
with two features of \ion{Si}{2} being especially distinct; (5) Weak
lines from \ion{S}{2}, and \ion{Si}{3} are also polarized, at a level
of about 0.3\%, but share the same polarization position angles as
those of the other polarized lines; (6) the high-velocity shell of
\ion{Ca}{2} has a velocity comparable to the blue wings of the strong
lines, but with a polarization amplitude and angle comparable to
\ion{S}{2}, and \ion{Si}{3}.

\section{The Inferred Ejecta Geometry}

\subsection{Bubbles and Plumes}

A critical issue is whether the observed polarization indicates
a clumpy ejecta or a globally aspherical structure with little
in the way of small scale features. The Q-U diagram provides a 
powerful tool to study the clumpiness of the ejecta. 
If the asymmetry is caused by an axially symmetric structure, the
resulting polarization will enscribe a straight line on the 
Q-U diagram. This is easy 
to understand as the polarization position angle is only related to
the position angle of the symmetry axis on the sky. Deviations from
the straight line on the Q-U diagram are expected only if the 
structure departs from axial symmetry, for instance by being 
fragmented and clumpy.

The observed polarization for each line can be fit by a linear relation:
$$U\ = \alpha+\beta Q. \eqno(1)$$
Such a fit gives the dominant axis of that particular line in 
the Q-U plane. The goodness of this fit is a direct measure of the 
smoothness of the ejecta.  For the strongest lines, the region for 
this linear fit was chosen to be from -25,000 \kms\ to -10,000 \kms\ for 
\ion{Si}{2} 635.5nm, 385.9nm, and \ion{Mg}{2}448.1 nm; 
this corresponds closely to the absorption minima of the 
P-Cygni profiles. The corresponding range was chosen to be
from -25,000 \kms\ to -13,000 \kms\ for the \ion{O}{1} 777.4 nm line to avoid
contamination by telluric features.
The situation is less clear for weaker lines which are often blended with
other lines. For the \ion{S}{2} ``W'' feature, the velocity range was taken to be 
from -15,500 \kms\ to -9,500 \kms and both lines were used for the linear fit.
For \ion{Ca}{2} IR triplet, the lines are blended and the range of the fit was 
taken to be 786.0 nm to 834.2 nm. 

\begin{figure}
\figurenum{6}
\epsscale{1.0}
\plotone{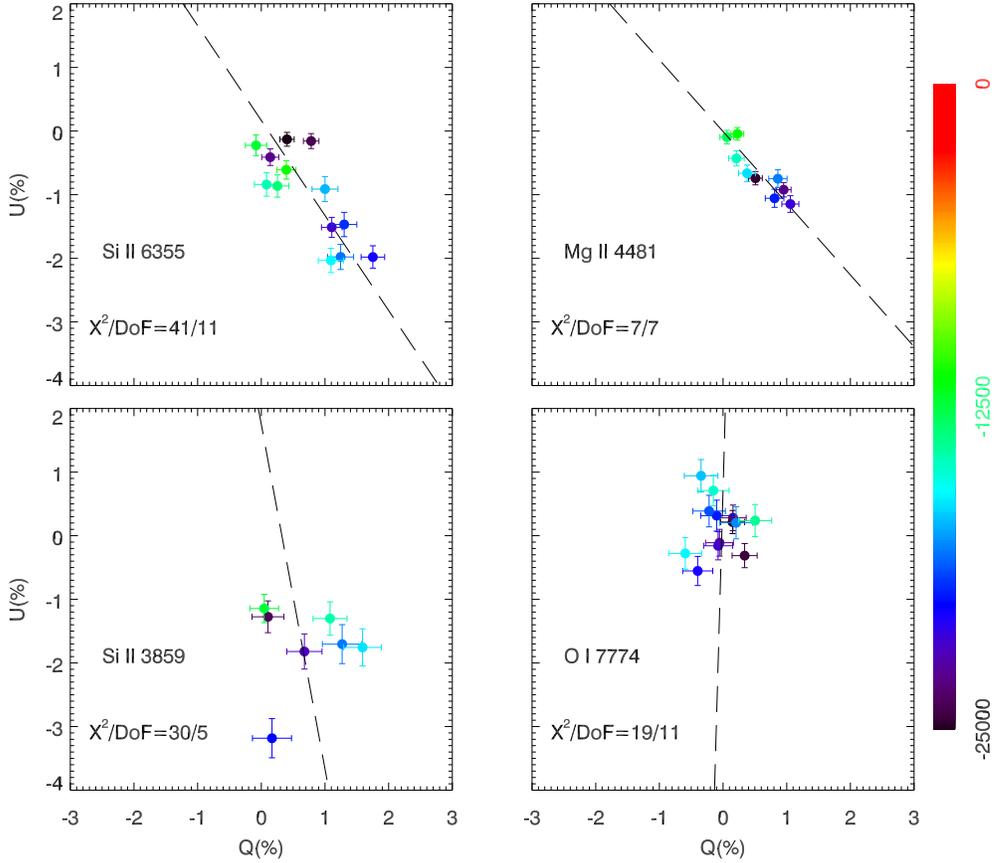}
\caption{
Q-U diagrams of the strongest lines and the corresponding linear fits. 
The dashed lines are the results of the linear fits. The ranges of the 
data points used for the fits are given in Table~2 and are from 
-25,000 \kms\ to -10,000 \kms\ for these lines. 
Clockwise from top left, the panels are for the lines of 
\ion{Si}{2} 635.5 nm (a, top left), \ion{Mg}{2} 448.1 nm (b, top right), 
\ion{O}{1} 777.4 nm (c, lower right), 
and \ion{Si}{2} 385.9 nm (c, lower left). Only the \ion{Mg}{2} line is 
consistent with a straight line, indicating a simple axially-symmetric
geometry with no detectable clumping (see text).
}
\label{quoflines}
\end{figure}

Fig.~\ref{quoflines} shows the Q-U diagrams for the strongest lines, 
\ion{Si}{2} 385.9 nm, 635.5 nm, \ion{Mg}{2}448.1 nm,
and \ion{O}{1} 777.4 nm. Table~2 gives the details of the
linear fits for all the lines that allow for a linear fit on the Q-U plane.
The columns in Table~2 are: the atomic line, the blueshift range 
of the data used in the fit, $\alpha$, $\beta$, the resulting position angle of the
straight line, the $\chi^2$, ,he degrees of freedom of the fit,
and the probability that the association with a straight line is by chance.

The linear fit provides a good description to the
observed Q-U diagram only for the \ion{Mg}{2} and the \ion{S}{2} lines. 
For the other ionic lines, the fits are highly 
inconsistent with a straight line. It is interesting to note that 
the \ion{O}{1} line, despite its lack of a well-defined axis, is 
inconsistent with a random Gaussian distribution and therefore the data
actually suggest intrinsic polarization of the oxygen, but  
with no well-defined symmetry axis. The \ion{Si}{2} and \ion{Ca}{2} lines 
all show clear evidence of a dominant axis, but with significant 
intrinsic deviations from the dominant axis. 
Only the \ion{Mg}{2} gives a good fit ($\chi^2$ per degree of freedom $\sim1$) 
indicating a distribution along a single, well-defined axis with
no measureable deviation due to fragmentation or clumping. The
\ion{Si}{2} 635.5 nm line gives a fit with $\chi^2$ per degree of freedom $~4$,
and a position angle that only differs from that of \ion{Mg}{2} by
about $2\sigma$. This suggests that the silicon may share the same
dominant axis as the magnesium, but with significant clumping.  

This analysis suggests that the distribution of silicon follows some
large scale asymmetry that gives rise to the dominant axis, 
but that there are also prominent smaller scale structures in
silicon that lead to significant deviations from the dominant axis 
in the Q-U diagrams.  The polarization of the \ion{Si}{3} line at 
lower velocity but with the same position angle suggests that this 
asymmetric silicon structure extends from at least the photosphere
at about 11,000 \kms\ all the way to the highest observable velocity 
for the \ion{Si}{2} lines, {\it i. e.}, 25,000 \kms. The transverse 
scale of this silicon-rich structure should be comparable to the size 
of the ejecta where the Thompson optical depth is around 1, so that the
asymmetric line opacity of the structure will have a large enough 
area covering factor to produce both the strong absorption and the 
large degree of polarization. This radially-extended silicon
structure is fragmented or clumpy. The size of the small scale 
structures that give rise to the departure from the dominant axis 
must be larger than 1,200 \kms, which corresponds to the approximate 
resolution of the data, and to the bin-size of the data points on the 
Q-U diagrams given in Fig.~\ref{quoflines}. Our data are not sensitive to
clumps smaller than 1,200 \kms.

The fact that the \ion{Mg}{2} line is {\it consistent} with a straight-line 
fit in the Q-U plane with no significant deviation from the principal 
axis means that, although its geometry follows quite nearly that of the 
\ion{Si}{2} globally, the magnesium has a smooth large structure with 
no noticeable small scale clumps or fragments. This has several 
implications. First, the success of the straight line fit for this line 
implies that the errors associated with the $Q$ and $U$ vectors are 
reasonable; this lends confidence in the robustness of the fits for 
the other lines. Secondly, this implies that the distribution of 
magnesium is less clumpy than that of silicon.  This could be 
understood if some amount of magnesium coexists with the silicon-rich 
clumps but that there are magnesium-rich regions that do not contain 
noticeable amount of silicon. The number density contrast of magnesium 
in different clumpy regions (presumably generated by explosive carbon 
burning) must be small; 
the number density contrast of silicon
in those different clumpy regions must be large.

The lack of a clear dominant axis for the \ion{O}{1} line and the 
obvious inconsistency of the data with random Gaussian distribution
on the Q-U plane suggests that the distribution of oxygen is even more
pervasive than magnesium, but the opacity must be very clumpy,
perhaps due to large variations of oxygen abundances in different 
chemical regions. 

The Q-U diagram shows undoubtably that at the highest velocity, 
the ejecta of SN~2004dt are highly asymmetric, clumpy and fragmented.

\subsection{Nucleosynthesis at the Highest Velocities}

The null detection 
of polarization across the \ion{O}{1} 777.4 nm line 
demands further explanation. From the spectral profile, we know that
oxygen co-exists with silicon, magnesium and calcium. 
Why does not the \ion{O}{1} 777.4 nm line share the same polarization
properties of lines from the other elements?

A simple explanation is that oxygen is distributed in a different way
from silicon, magnesium and calcium, 
although they share the same velocity space. Oxygen is likely more
pervasive in the outer region, but the protrusions of 
products from nuclear burning are distributed in holes or filaments 
in this region. Such protrusions also contain oxygen. In such a 
geometry, the polarization from oxygen and from the other elements are 
drastically different as their geometrical structures are not
identical but occupy different regions.

If so, then the null detection of polarization across the 
\ion{O}{1} 777.4 nm line implies that the oxygen rich region does
not exhibit the same asymmetric geometry as 
the magnesium/silicon rich regions. This would 
require that the magnesium/silicon rich regions also contain a 
significant amount of oxygen so that oxygen appears to be more
``spherically'' distributed. Indeed, in explosive carbon burning,  
oxygen is expected to co-exist with magnesium, a principal 
burning product. 

The observed high degree of polarization of high-velocity magnesium 
suggests that the magnesium-rich structure was created by explosive
carbon burning. The fact that the magnesium structure is polarized 
with about the same amplitude and position angle as the silicon 
lines also shows that magnesium is not confined to magnesium-rich,
silicon-poor regions, but also exists in silicon-rich regions. This 
in turn suggests that the oxygen that must be present outside the 
magnesium/silicon structure in order to have a very small net 
polarization must be the pre-explosion oxygen from the progenitor 
white dwarf.  

The unusual polarization of SN~2004dt is closely 
related to the observed velocity structures of the various ions. 
In the outermost portions, pre-explosion oxygen co-exists with
newly-synthesized Si, Mg, and Ca. This outermost layer moves 
at a velocity around 17,000 \kms\ and the highest velocity approaches 
25,000 \kms. The silicon-rich layer extends deep inwards to a velocity of
around 11,000 \kms, as evidenced by the presence of \ion{Si}{3} and 
\ion{S}{2} lines at that velocity. The broadness and unusually high 
blueshift of the \ion{Si}{2} lines are evidence that these lines are 
optically thick.  Iron and sulfur are noticeably absent in the high-velocity 
layer.  The strong polarization observed across the P-Cygni lines of the 
newly-synthesized elements is the result of the 
asymmetric distribution of the chemical elements in this this layer. 
The polarization position angles of these lines are similar, suggesting
that they share a common geometrical structure; but with oxygen permeating
the whole structure, magnesium occupying a globally asymmetric region, and
silicon occupying a similar globally asymmetric region but, in addition,
being broken into clumps or fragments.

The structure of the ejecta is sufficiently asymmetric that spectral 
lines can be formed in significantly different velocity regions.  
The attribution of the polarization to asymmetric chemical distributions 
requires the structures to have considerable radial extent because 
the angle of the dominant axis is the same for lines formed at different 
depths, that is, for both \ion{Si}{2} at 17,000 \kms\ and \ion{Si}{3} at 
11,000 \kms. The photosphere in the conventional sense of Thompson optical 
thickness probably has a complicated geometry that will complicate 
attempts to understand the radiative processes. 
For simplicity, one may still assume a 
photosphere that is located at a velocity much smaller than the 
17,000 \kms\ as derived from the absorption minima of the P-cygni 
lines of \ion{Si}{2} lines. The weaker \ion{S}{2} ``W" lines and 
\ion{Si}{3} lines at 439.5 454.0, and 464.0 nm may provide a hint of the 
location of the photosphere. If so, at least along certain lines of 
sight the photosphere is at a velocity as low as 11,000 \kms. This 
rather low photospheric velocity may not be as surprising. 
From light curve fitting, we know that the pre-maximum luminosity of 
SN~Ia rises approximately as $t^{2}$ (Goldhaber et al. 2001), 
which implies a photosphere at a fixed velocity if the atmosphere 
is approximated by a blackbody. A photospheric velocity of 11,000 \kms\  
agrees with the observed luminosity and light curve behavior of 
typical SN~Ia.  

\subsection{Continuum polarization and the geometry of the photosphere}

We note that most of the discussion in this paper relies on 
polarization variations across spectral features, not on the
level of continuum polarization. We will not 
elaborate on the geometrical structure of the photosphere.
With the complicated geometry shown by the various lines,
the continuum may be formed in a complicated clumpy environment.
It is likely not to have a smooth geometry like the one assumed for 
ellipsoidal models. 
The fact that the emission peaks of the strongest P-Cygni
profiles are all blueshifted by about 4,000 \kms\ (see Fig.~ref{lineprofies})
suggests that there is a substantial amount of electron
scattering at high velocities.
The rather extended region of electron scattering effectively
blocks the receding side of the ejecta and can shift the 
emission peak of the P-Cygni profile to the blue.
This interpretation of the strong blueshift has to be 
reconciled with the presence of faint spectral line features 
at velocities as low as 11,000 \kms\ through more detailed line
profile modeling. Lacking such models, it should be remarked, however, 
that a clumpy ejecta is able to produce P-Cygni profiles with
a blueshifted emission peak together with absorption features at 
low velocity (Wang \& Hu 1994). In such a scenerio, the patchy opacity 
structure allows formation of spectral features at low 
velocity together with shielding of the receeding side of the
supernova atmosphere.

Taking the adopted components of interstellar polarization at face value, we
can see that the level of continuum polarization can be as low as 0.2-0.3\%,
consistent with what was found for SN~2001el. 
The amplitude of continuum polarization is larger in the red part of the spectrum
than in the blue part, but with the sign of the polarization flipped. 
The relatively low level of continuum polarization does suggest that the asphericity of 
the photosphere is not very large, perhaps only at a level of 
less than 10 percent as has been derived for other Type Ia supernovae such 
as SN~1996X (Wang, H\"oflich, Wheeler 1997) and SN~2001el (Wang et al. 2003a).
The exact level of continuum polarization awaits a full analysis of the data set
which covers post-maximum phases of the supernovae. 

\section{SN~2004dt as a member of a sub-class of SN~Ia}

\subsection{Spectroscopically similar events}

\begin{figure}
\figurenum{7}
\epsscale{.8}
\includegraphics[angle=90,scale=0.75]{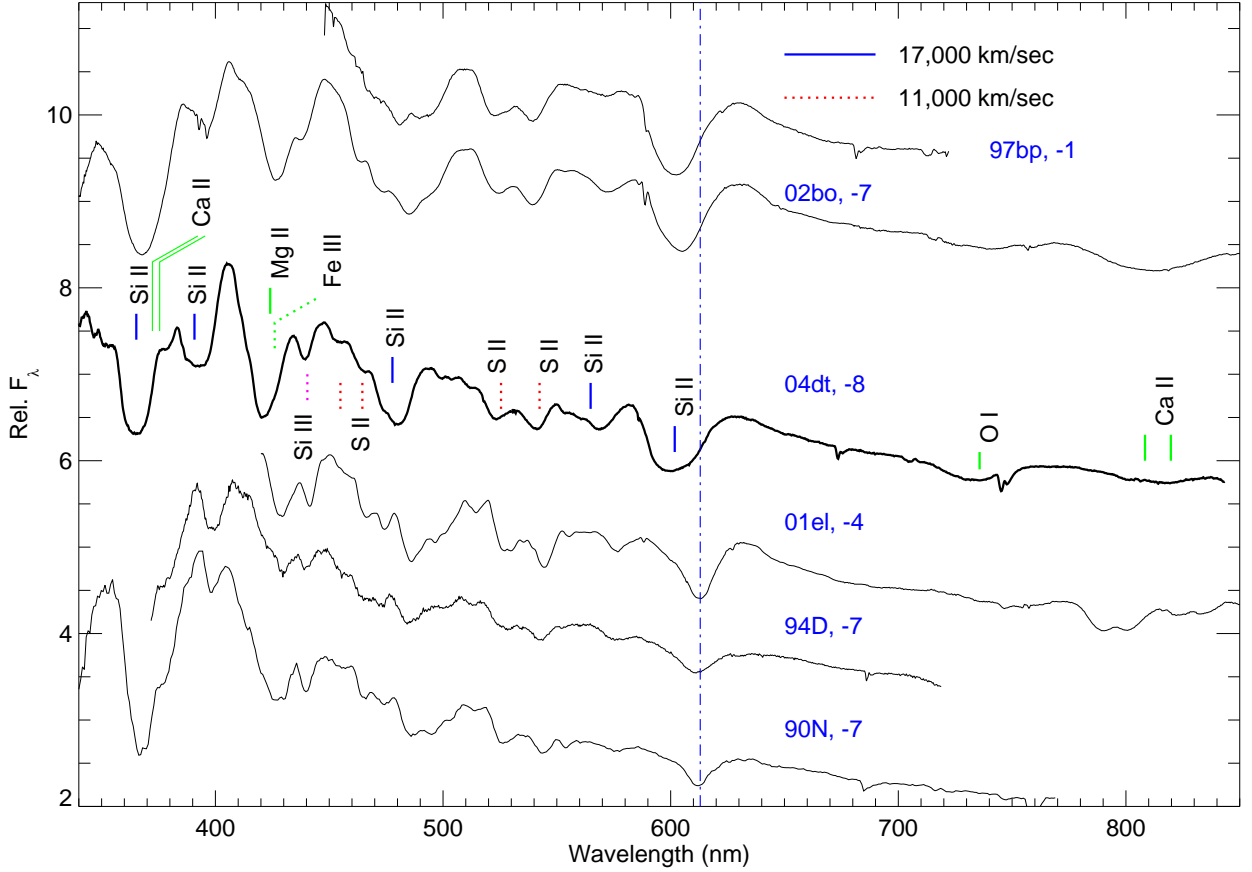}
\caption{Comparisons of spectra of SN~1997bp, SN~2002bo, SN~2004dt, SN ~1990N, SN~1994D, and 
SN~2001el before optical maximum. The data of SN~1997bp were obtained at the 2.1 meter telescope
of McDonald Observatory through our program of supernova polarimetry (Wang \& Wheeler 1997). 
The data on SN~2002bo and SN~2001el were obtained 
by our supernova polarimetry program at the VLT (Wang et al. 2003a).
The data of SN~1994D is from Patat et al. (1996). The data of SN~1990N
is from Jeffery et al. (1992).
The \ion{Si}{2} 635.5 nm lines
of SN~1997bp, SN~2002bo, and SN~2004dt are significantly broader than 
those of SN~1990N, SN~1994D, and
SN~2001el around optical maximum. The vertical line marks the absorption 
minimum of the \ion{Si}{2} 635.5nm line of SN~2001el, SN~1994D, and 
SN~1990N. The \ion{Si}{2} 635.5nm lines of SN~1997bp, SN~2002bo, 
SN~2004dt show significantly larger blue shift than those of 
SN~2001el, SN~1994D, and
SN~1990N.
}
\label{othersn}
\end{figure}

SN~1997bp and SN~2002bo displayed high velocity 
absorption features similar to those of SN~2004dt. 
Strong polarized features were also observed 
across the \ion{Si}{2} 635.5 nm line for both SN~1997bp and SN~2002bo. 
Discussion of the polarization
features of SN~1997bp and SN~2002bo will be given in a forthcoming paper.
We discuss here only their distinct spectroscopic behavior. As pointed in
Benetti et al. (2004), supernovae such as SN~1984A (e.g. Branch 1987) 
and SN~2002bo show unusually high velocity lines. A 
comparison of the spectra of SN~1997bp, 
SN~2002bo, SN 2004dt, and some spectroscopically normal SN~Ia is shown 
in Fig.~\ref{othersn}.   
The characteristic behavior of SN~1997bp, SN~2002bo, and SN~2004dt
is that they appear spectroscopically normal, but with the lines 
much broader and at much higher velocities. 
Benetti et al. (2004) suggest that for this group of supernovae the 
burning to silicon penetrated to higher layers than in more normal SN~Ia.
The high velocity, broad lines can be understood by the presence 
of a fast moving shell which is rich in intermediate mass elements 
and pre-explosion oxygen. The extension of this shell in velocity
space requires detailed spectral modeling.

Both SN~1997bp, and SN~2002bo show strong narrow \ion{Na}{1} D absorption 
that implies that they suffer considerable 
amount of dust extinction from the host galaxy. The equivalent widths of
the \ion{Na}{1} D absorption are 1.5 \AA, and 2.5 \AA\ for SN~1997bp and SN~2002bo,
respectively. No strong Na I D line is detected for SN~2004dt. 
For SN~1984A, Barbon et al. (1989) derived $E(B-V)\ = \ 0.45\pm0.05$. 
Kimeridze \& Tsvetkov (1986) found $A_B\ = \ 1.2\pm0.2$.
Although these values are uncertain, the observed color around maximum does
suggest large interstellar extinction. For SN~1997bp, 
Riess et al. (1998) give $A_V\ = \ 0.62$, 
suggesting $E(B-V)\ \approx \ 0.2$. A significant amount of dust 
along the light of sight of SN~1997bp in the host galaxy is also 
consistent with the presence of a strong \ion{Na}{1} D line.
Benetti et al. (2004) found $E(B-V) \ = \ 0.43\pm0.10$ for SN~2002bo,
in agreement with the observed strong \ion{Na}{1} D in SN~2002bo.
SN~2004dt does not show a noticeable \ion{Na}{1} D line, and 
final light curves are not yet available to estimate the dust 
content on the line of sight; however, our flux-calibrated spectrum shows 
$B-V\ \approx\  0.07$ at around day -7. Although uncertain, this 
implies an $E(B-V)$ of 0.14-0.2
when compared to colors of typical SN~Ia at comparable epoch. 
Typical $B-V$ for an unextinguished supernova at this phase is around -0.07
({\it e.g.} Phillips et al. 1999).
These values of $E(B-V)$ are significantly larger than typical values of
$E(B-V)$ found for Type Ia supernovae (Knop et al. 2003).

It is not clear to us whether the extinction of these supernovae
is caused by circumstellar or interstellar dust. High resolution spectroscopy
may be helpful in this regard, but it may be difficult to 
use high resolution spectroscopy alone to tell whether the dust
is within a few parsecs or a few hundred parsecs from the
progenitor. Late time light echo observations (Patat 2004)
combined with late time polarimetry may set more direct constraints
on the location of the absorbing dust.
A larger sample of SN~2004dt-like objects will resolve 
the issue of how strongly they are associated with dusty environments.

\subsection{SN~20024dt and other polarized supernovae}

SN~2004dt continues to add to the data base that strongly 
suggests that SN~Ia are polarized in a very interesting fashion, but
with a variation from event to event that we are just 
beginning to probe.

The observations have revealed
both high and low velocity spectral features about a week 
before optical maximum. Strong variation of
polarization across spectral lines of intermediate mass elements, 
especially silicon, is observed; but most puzzling is non-detection 
of polarization in the \ion{O}{1} 777.4 nm line.

SN~2004dt is different than SN~2001el. For the
latter event, the strongest polarization feature was the
high-velocity component of the Ca II IR triplet with \ion{Si}{2}
polarized at a much lower level than the Ca II IR triplet. For SN~2004dt, the
polarization of the Ca II IR features is prominent, but  
the polarization levels of the lines of \ion{Si}{2} are the strongest. 
SN~2001el showed continuum polarization at the level of about 0.3\%. 
SN~1996X tentatively showed similar levels of continuum polarization.
As we discussed before, the precise level of continuum polarization 
of SN~2004dt is difficult to determine because of the highly-polarized 
spectral features. With the adopted level of interstellar polarization, the
continuum level may be comparable to that of SN~2001el and SN~1996X; 
however, the large polarization across most of the spectral lines
makes SN~2004dt distinctively different from SN~2001el.
In Wang et al. (2003a), we argued that it is important to study 
the correlations among the degree of polarization of the 
high-velocity \ion{Ca}{2} IR triplet, the strength of the 
high-velocity component of the \ion{Ca}{2} IR triplet, and the velocity
of the \ion{Ca}{2} IR triplet. The observations of SN~2004dt are
consistent with a positive correlation
between the strength of the high-velocity \ion{Ca}{2} IR triplet 
and the degree of polarization. The observations also seem to
confirm the speculation that higher velocity events should show
larger polarization. Obviously, more spectral polarization data are needed.

\subsection{SN~2004dt and Sub-Types of SN~Ia}

All of the four Type Ia supernovae identified to have abnormally high 
velocity \ion{Si}{2} lines show also evidence of unusually large 
reddening by the dust in the host galaxy. 
Both SN~1997bp and SN~2002bo are found to be significantly
polarized, at levels around 1-2\%. There were no polarimetry data on SN~1984A.
This group of objects is likely to be characterized by
high velocity, high dust extinction, and intrinsically large polarization before
optical maximum.

We do not yet know whether these high velocity Type Ia supernovae 
form a separate sub-group of SN~Ia, or whether they are simply a familiar 
sub-group of objects viewed at a special angle.
For the latter possibility, SN~1991T-like events may be candidates
to be the counterparts viewed at different angles. The only other 
sub-group of SN~Ia that is
known to be associated with dusty environments are SN~1991T-like events.
Perhaps SN~1991T-like objects are viewed in a direction with
very little intermediate mass elements in the outer layers, whereas
SN~2004dt-like objects are the same objects viewed in the direction with
substantial pre-explosion elements on the outside.

\section{Comments on theoretical models of Type Ia supernova explosions}

SN~Ia involve the combustion of a degenerate C/O white dwarf of 
the Chandrasekhar mass in a binary system (Whelan \& Iben 1973; 
Nomoto 1980, 1982).  Major theoretical questions concern the 
ignition and propagation of the flame through the white dwarf. 
Three major scenarios have been proposed: 
detonation (Arnett 1969; Hansen \& Wheeler 1969), deflagration 
(Nomoto, Sugimoto, \& Neo 1976; Nomoto, Thielemann, \& Yokoi 1984), 
and delayed detonations (Khokhlov 1991; Yamaoka et al. 1992; 
Woosley \& Weaver 1994). Three-dimensional calculations show
that pure deflagration models leave unburned carbon and oxygen 
mixed into the center (Gamezo, Khokhlov \& Oran 2002; 
Reinecke, Hillebrandt, Niemeyer, 2002). 
Delayed-detonation models leave very little unburned carbon and 
tend to produce the observed stratified composition structure
(Gamezo, Khokhlov \& Oran 2004; Marion et al. 2003; 
H\"oflich et al. 2001, 1995, Wheeler et al. 1998).

\subsection{Deflagration}

In SN~2004dt, we see a highly asymmetric and clumpy structure 
in silicon that extends from at least as deep as 11,000 \kms\ to
at least 17,000 \kms.  This fragmented structure could, in principle, 
be consistent with the turbulent structure produced in 3-D simulations 
of deflagration models (Gamezo et al. 2002; R\"opke, Niemeyer, J. C., 
\& Hillebrandt 2003).  The layer at velocity around 17,000 \kms\ shows
pre-explosion oxygen and fragments of intermediate mass elements, but with
no iron. Deeper inside, sulfur is found at a velocity around 
11,000 \kms\ which shares the same asymmetry signature
as the high velocity, oxygen-rich layer.

In deflagration models,
silicon co-exists with iron; the absence of strong iron features 
in the spectrum may be inconsistent with published 
pure deflagration models, but we caution that the exact 
amount of iron that can be hidden in the spectra without 
producing pronounced spectral features is uncertain and quantitative 
spectral modeling is needed to set the constraints on deflagration
models. The velocities of the silicon and magnesium
are significantly higher than predicted by pure deflagration 
models. It thus seems that turbulent deflagration models are capable of
producing the observed clumpy composition structure, but are perhaps 
incapable of producing the kinematics of the ejecta. 
Deflagration models also predict unburned matter, C and O, mixed
down to low velocities. Subsequent observations of this event
will help to put constraints on this possibility. Deflagration
models also tend to leave a layer of completely unburned C
and O on the outside. It is very important to put limits on
the amount of carbon in the spectra. 

\subsection{Delayed-detonation}

An alternative is the delayed-detonation (DD) model (Khokhlov 1991).
This model assumes that at a certain stage of deflagration, the nuclear
burning becomes supersonic and therefore switches to a detonation.
In order to have a high velocity partially burned layer, 
there must be significant pre-expansion 
(due to the subsonic deflagration stage) 
before the onset of detonation, which sufficiently 
reduces the density of the progenitor white dwarf before the 
nuclear burning flame propagates to the surface layers.
It may also be possible that there is a layer on  
the surface of the progenitor white dwarf with a density
that is too low to burn at all.
Because of the low density near the surface,
the products of the nuclear reactions are mostly intermediate mass 
elements. 
A strong detonation wave is unlikely to generate a turbulent silicon 
layer such as we apparently observe in SN~2004dt. 
One possible way out is that the
detonation wave weakens significantly near the surface of the star,
in which case the cellular structure associated with 
detonation instability (Boisseau, et al. 1996; Gamezo et al. 1999) 
can grow to significant scales.

Alternatively, the detonation could be switched on at a stage where the
deflagration flame has propagated sufficiently close to the low density part
of the expanding white dwarf that the products of detonation bear the
imprint of the complex structure of deflagration.

A delayed detonation would be able to boost
the outer layers to the velocities we observe. It may also be consistent with
the absence of iron features at very high velocity. Detailed 
modeling is needed in order to understand the clumpy structure of the 
products of nuclear burning, especially silicon, that this data
indicates.

\subsection{Off-centered delayed-detonation}

A different model which naturally produces a globally aspherical
Type Ia is given in Livne (1999) and more recently
in 3-D calculations by Gamezo et al. (2004) in which the detonation
is triggered at a point off the center of the white dwarf
rather than in a spherical shell. The result is that the
inner nickel region is offset from the centroid of the density
distribution. The resulting off-center production of gamma-ray
energy input can induce polarization even in a spherical
density distribution (Chugai 1992; H\"oflich 1995). In addition, the
silicon is predicted to be left with a distribution that is
displaced from the centroid of the density distribution. This
leads naturally to the possibility of a globally asymmetric 
silicon distribution that has the potential to produce significant
polarization by large scale blocking of the photosphere as
well as smaller scale clumps of silicon resulting from the
turbulent burning. This sort of model thus has the
capability of producing all the polarized signatures of the silicon-rich layer.
Since the photosphere might also be significantly aspherical
from the energy input or the asymmetric density distribution, a
large continuum polarization may be produced by such a model; this
should be confronted with detailed theoretical modeling. 
This model can be tested by spectroscopic observations in the nebular phase. 
We will defer a detailed discussion of this model to a later paper when
the analysis of the complete data set of our observations is 
undertaken.

\subsection{Gravitationally-confined detonation}

In the model recently
proposed by Plewa, Calder \& Lamb (2004), 
the explosion begins as a deflagration born slightly off-center
in the white dwarf. Buoyancy dominates the evolution of the 
deflagration, resulting in a very rapidly rising plume of
burned material (see also Livne, Asida \& H\"oflich 2005). 
This plume displaces the outer (unburned) layers
of the star, which propagate ahead of burned bubble material across
the surface of the star. This flow converges at the point opposite
from where the plume emerges, and the resulting rapid compression
of the unburned material triggers a detonation. 
An advantage of this model is that it 
may not require triggering the detonation artificially.  This model 
has about 1 percent of a solar mass of iron group elements 
on the surface of the white dwarf resulting from the surface 
flow of the plume material. As for pure deflagration models, detailed 
spectroscopic modeling is needed to tell if this amount of iron is 
allowed by observations.  In addition, the requirement of 
significant pre-expansion so that sufficient intermediate
mass elements are produced in the velocity range from 11,000 \kms\ to
25,000 \kms\ to match the observed spectra may further constrain this model. 
Note that pre-expansion is necessary to produce intermediate mass
elements in the outer layers, but can significantly weaken the shock of the
merging plume matter; this will work against the triggering of the detonation.
Therefore these observations will set limits on the parameter space
where this mechanism may work. Due to pre-expansion at the stage of
detonation, it is also possible that this mechanism will 
produce highly aspherical chemical structures such that one
side (where detonation is triggered) consists of mainly iron 
rich material on the outside, whereas the other side (where
the burned bubble emerges) consists of mainly of intermediate 
mass elements. Depending on which side faces the observer,
the target might appear as an iron-rich SN~1991T-like, or a 
silicon-rich SN~2004dt-like; however, the details of these 
constraints await thorough theoretical models of these data.

\section{Implications for Supernova Cosmology}

This study illustrates that the nature of Type Ia supernovae is more
heterogeneous than is easily revealed by observing methods 
that are less telescope time intensive than spectropolarimetry.  
We have shown that polarimetry provides information that is often 
strongly complimentary to other observables and so can provide 
important discrimints among theoretical models of SN Ia explosions.

In order to make SNe Ia better understood standard candles it will be
essential (i) to obtain enough data to extract the core commonalities
from all the individual cases and (ii) to find ways of inferring
certain special conditions also from low-resolution spectra and
photometry because only they permit SNe to be observed at cosmological
distances.
If unrecognized sub-types of SNe occur in different proportions at
different redshifts or in different environments, supernova cosmology
experiments could be severely compromised by such uncertainties.

There are two effects of asymmetry on the observed magnitudes of a
supernova. First, asymmetry affects the total integrated light
from a supernova in a particular filter, therefore causing a
view-angle dependence of distances. Secondly, asymmetry will cause
intrinsic color dispersions which can make the determination
of color-excess due to dust in the host galaxy difficult. The second
effect can become very important as errors on color
are amplified when extinction corrections are to be applied.

The
nature of events like SN~2004dt may require this group to be separated
from other supernovae, and a re-calibration of the methods
(e.g.\ Phillips et al. 1999; Wang et al. 2003b) commonly used to derive
distances and dust reddening of SN~Ia. This supernova may represent a 
particular sub-set of SN~Ia, along with SN~1984A, SN~1997bp, and SN~2002bo, 
with especially high-velocity matter for which the explosion mechanism is
significantly different than that for their lower velocity
counterparts such as SN~1994D.

An association with dusty environments would suggest also that these
events are likely to be from younger progenitor stars, and that they
may be more abundant at higher red shifts than in our local
universe. Identifying these events and applying the proper relations
for distance calibration is also critical for removing the effects of
progenitor evolution.

\section{Summary and Discussion}

Spectropolarimetry is a straight forward, powerful diagnostic
tool for supernova explosions. Contrary to conventional perceptions
that the interpretation of polarimetry data is complicated and
heavily model dependent, many important insights on the geometric
structure of supernovae ejecta can be derived from diagnostic analyses 
that do not rely on detailed modeling.

In this study, we have presented high quality spectropolarimetry of
the Type Ia SN~2004dt. These data allow us to probe the geometric
structure of the ejecta from around 11,000 \kms\ all the way to 
about 25,000 \kms.

SN~2004dt adds to the examples of how these simple 
diagnostic tools can be applied to well observed supernovae to 
study the geometric structure of the ejecta. 
Spectroscopically, we found that most of the spectral features 
can be identified with intermediate mass elements with 
little or no contributions from iron. From spectropolarimetry, 
we found the high velocity regions of SN~2004dt to be
highly aspherical, clumpy and probably turbulent.

The results of this study provide significant constraints on 
theoretical models of Type Ia explosions.  Polarimetry of SN Ia 
probes the origins of the intrinsic diversity of SN~Ia, and 
is important for improving them as precise tools of cosmology.

{\bf Acknowledgments:}
The authors are grateful to the European Southern Observatory
for the generous allocation of observing time. They especially thank 
the staff of the Paranal Observatory for their competent and never-tiring 
support of this project in service mode. We are grateful to Weidong Li for
sharing photometric data before publications, and to Tomesz Plewa for
helpful discussions. The research of LW is supported in part by 
the Director, Office of Science, Office of High Energy and Nuclear Physics,
of the U.S. Department of Energy under Contract No. DE-AC03-76SF000098.
This work is performed in part during a visit of LW
to Purple Mountain Observatory, Nanjing, China. The research of JCW 
is supported in part by NSF grant AST-0406740.

\newpage

\begin{table}[h]
\caption{Summary of Observed Polarization Features}
\begin{tabular}{c|c|c c|c c}
\hline
\hline
Ion & Rest wavelength(s) [nm] & \multicolumn{2}{c|}{Radial velocity [\kms]} &
\multicolumn{2}{c}{Polarization} \\
           &                 &   bulk  &  blue edge  &  degree [\%] &
angle \\ 
\hline
Cont   & 350-600           & 11,000 &  N/A   & 0.0    & N/A \\
Cont   & 600-850          & 11,000 &  N/A   & -0.3    &  60\degree \\
\hline
\ion{S}{2}  & 471.6, 481.6, 543.3, 545.9 & 11,000 & 15,000  & 0.3 & 150\degree \\
\ion{Si}{3} & 455.3/456.8/457.5 & 11,000 &        &  0.3    & 150\degree \\
\hline
\ion{Ca}{2} & 850.0/854.2/866.2 & 16,000 & 27,500 & 0.7-2   & 150\degree \\
\ion{Mg}{2} & 447.1            & 17,000 & 28,000 &    1    & 150\degree \\
\ion{O}{1}  & 777.4            & 17,000 & 27,500 &  $<$0.2 & N/A \\
\ion{Si}{2} & 385.0, 413.0, 504.1, 505.6 635.5 & 15,000-17,000 & 25,000 & 0.5-2 & 150\degree
\\
\hline
\end{tabular}
\end{table}

\newpage

\begin{table}[h]
\caption{Properties of Polarized Lines}
\begin{tabular}{l|r|rr|l|rrr}
\hline
\hline
Line & V$_{range}$ (\kms) & $\alpha$ & $\beta$ & P.A.\tablenotemark{b} & $\chi^2$ & DoF & Prob. \\
\hline
\ion{Si}{2} 635.5 & (-25,000, -10,000)& 0.17$\pm$0.15   & -1.49$\pm$0.19  & 151.9\degree$_{-1.6}^{+1.9}$ & 41 & 11 & 1.8 10$^{-5}$\\
\ion{Si}{2} 385.9 & (-25,000, -10,000)& 1.75$\pm$4.28   & -5.42$\pm$15.1  & 140.2\degree$_{-3.8}^{+81.8}$& 30 & 5  & 1.1 10$^{-5}$\\
\ion{Mg}{2} 448.1 & (-25,000, -10,000) &  -0.01$\pm$0.12 & -1.13$\pm$0.18  & 155.8\degree$_{-2.1}^{+2.6}$ & 7 & 7 & 0.34\\
\ion{O}{1} 777.4 & (-25,000, -13,000) & 0.72$\pm$0.66   &  35.4$\pm$293.8  & 44.2\degree$_{-89.1}^{+0.7}$ & 19 & 11 & 0.05\\
\ion{Si}{3} 456.5 & (-15,000, -5,000) & 0.09$\pm$0.46   & -1.22$\pm$1.02  & 154.6\degree$_{-7.6}^{+19.6}$ &11 & 4 & 0.02\\
\ion{S}{2} 562.3,544.8 & (-15,000, -9,500) & -0.24$\pm$0.18 & -0.96$\pm$0.45 & 158.1\degree$_{-5.4}^{+8.4}$ & 9 & 6 & 0.15\\
\ion{Ca}{2} IR Triplet & (-24,000, -7,000)\tablenotemark{a} & 0.29$\pm$3.6 & -8.5$\pm$10.9 & 138.3\degree$_{-1.9}^{+75.0}$ & 38 & 17 & 0.0
04\\
\hline
continuum & (640, 700)\tablenotemark{c} & 0.11$\pm$0.04 & 1.1 $\pm$ 0.3 & 0.0\degree$_{-7.1}^{7.2}$ & 46 & 23 & 0.05 \\
\hline
\hline
\tablenotetext{a}{Velocity calculated with respect to the
restframe wavelength of 854.2 nm.}
\tablenotetext{b}{The errors on P. A. are highly non-Gaussian when the slopes 
of the linear fits are not well constrained. The P.A. is periodic with
period of 90\degree. The angle is practically undetermined for \ion{Si}{2} 385.9 nm, 
\ion{O}{1} 777.4 nm, and the \ion{Ca}{2} IR Triplet. }
\tablenotetext{c}{Wavelength range in nm}
\end{tabular}
\end{table}

{}

\newpage

\newpage

\newpage
\newpage

\newpage

\newpage

\newpage

\end{document}